# Increasing performance of planar PIV

Dinar Zaripov[1,2], Renfu Li[3], Mikhail Tokarev[2], Dmitriy Markovich[2]

[1] School of Mechanical Science and Engineering, Huazhong University of Science and Technology, 430074 Luoyu Road 1037, Wuhan, China
E-mail: zaripov.d.i@gmail.com

[2] Institute of Thermophysics SB RAS, Lavrentyev ave. 1, 630090 Novosibirsk, Russia

[3] School of Aerospace Engineering, Huazhong University of Science and Technology, Wuhan, China
E-mail: renfu.li@hust.edu.cn

**Abstract** To date, the iterative image deformation method of PIV for two-pulse measurements is widely used in experimental fluid dynamics due to its robustness in many scientific and industrial applications. However, it has a known limitation associated with the difficulty of image deformation due to errors that occur when calculating the necessary velocity derivatives using higher-order schemes. In this work, we propose a simple method that helps to noticeably improve the accuracy of the measured velocities and their derivatives, and thereby increase the spatial resolution. The method does not require the recovery of erroneous velocity vectors, avoids the numerical calculation of derivatives necessary for the interrogation window deformation using finite difference schemes, and can be easily applied in practice.

Over the past three decades, the particle image velocimetry (PIV) method has become a reliable tool for measuring various fluid flow quantities. Many approaches have been developed to expand the area of its application and increase the reliability of the data obtained (Raffel et al. 2018). Perhaps the most advanced method today is based on the iterative image deformation approach proposed by Huang et al. (1993) and substantially improved then by Scarano (2002) using the WIDIM (Window Displacement Iterative Multigrid) algorithm, which allows a significant reduction of the in-plane loss of pairs and increases the dynamic velocity range.

The principle of this method is that if a certain initial displacement field $s = s(s_x, s_y)$ is known, then when cross-correlating, one can use the interrogation windows (IWs) that are shifted and deformed according to this field. Usually, this deformation is carried out taking into account the terms of the $0^{th}$, $1^{st}$, or $2^{nd}$ orders of the accuracy of the Taylor series expansion of the displacement field relative to the IW center $\vec{x}_0 = \vec{x}_0(x_0, y_0)$:

$$s(\vec{x}) = \underbrace{s(\vec{x}_0)}_{\text{0th order}} + \underbrace{\sum_{k=1}^{2} s_k(\vec{x}_0)\Delta x_k}_{\text{1st order}} + \underbrace{\frac{1}{2!}\sum_{k,l=1}^{2} s_{kl}(\vec{x}_0)\Delta x_k \Delta x_l}_{\text{2nd order}} + \underbrace{\cdots}_{n\text{th order}} + \underbrace{e}_{\substack{\text{truncation}\\\text{error}}}, \qquad (1)$$

with $e = O[(\Delta x)^3, (\Delta y)^3]$, $\Delta x = x - x_0$, $\Delta y = y - y_0$. Thus, this method takes into account the non-uniformity of the velocity profile within IW, with more complex IW deformation when higher order of accuracy is taken into account in Eq. (1).

However, despite the wide popularity of this approach, it has a number of limitations, mainly associated with the difficulties in the numerical calculation of the velocity derivatives included in Eq. (1). The situation is complicated by the presence of a large number of outliers, especially at the boundaries of the measurement area. In this case, after each iteration, an additional step is required to reconstruct the missing vectors that practically appear after an outlier detection procedure, such as the universal median test (Westerweel and Scarano 2005). As a result, the velocity derivatives are affected not only by the measurement error but also by the interpolation procedure. Since the estimation of higher-order derivatives is more sensitive to such errors, the use of the iterative image deformation method in some cases may not lead to a decrease in the measurement error due to incorrect IWs deformation used at the next iterations. For this reason, in practice, they are most often limited to the $1^{st}$-order of accuracy, thereby neglecting the non-uniformity of the velocity profile within IW and, therefore, biasing the corresponding measured values and losing the spatial resolution.

To solve the problems stated, the following method is proposed. According to (Tiwari and Kuhnert 2001), using the least-squares method based on minimizing the weighted truncation error in Eq. (1), any variable $s = s(x, y)$ as well as its derivatives at any mesh point $i$ with coordinates $\vec{x}_i = \vec{x}_i(x_i, y_i)$ can be estimated using the values of $s$ at $p$ neighboring mesh points as follows:

$$a = \left(M^T W M\right)^{-1} \left(M^T W\right) b, \tag{2}$$

with

$$a = \left(\underbrace{s}_{\text{0th order}}, \underbrace{\frac{\partial s}{\partial x}, \frac{\partial s}{\partial y}}_{\text{1st order}}, \underbrace{\frac{\partial^2 s}{\partial x^2}, \frac{\partial^2 s}{\partial x \partial y}, \frac{\partial^2 s}{\partial y^2}}_{\text{2nd order}}, \underbrace{\cdots}_{\substack{n\text{th} \\ \text{order}}}\right)^T, \tag{3}$$

$$M = \begin{pmatrix} 1 & \Delta x_1 & \Delta y_1 & 0.5\Delta x_1^2 & \Delta x_1 \Delta y_1 & 0.5\Delta y_1^2 & \cdots \\ 1 & \Delta x_2 & \Delta y_2 & 0.5\Delta x_2^2 & \Delta x_2 \Delta y_2 & 0.5\Delta y_2^2 & \cdots \\ \vdots & \vdots & \vdots & \vdots & \vdots & \vdots & \vdots \\ 1 & \Delta x_p & \Delta y_p & 0.5\Delta x_p^2 & \Delta x_p \Delta y_p & 0.5\Delta y_p^2 & \cdots \end{pmatrix}, \tag{4}$$

$$b = (s_1, s_2, \ldots, s_p)^T, \tag{5}$$

$\Delta x_p = x_p - x_i$, $\Delta y_p = y_p - y_i$, $s_p$ is the known value of the variable $s$ at the $p^{th}$ neighboring mesh point, $W$ is the diagonal matrix of weights, the elements of which are determined in the present research as the inverse squares of the distances between the current and neighboring mesh points. Concerning PIV, the displacement fields of $s_x^{\kappa-1}(x, y)$ and $s_y^{\kappa-1}(x, y)$ found at the previous iteration $\kappa - 1$ can be used as the variable $s$. Thus, by solving the system of linear algebraic equations (2) for $s_x$ and $s_y$ separately, we can find both velocity components and all derivatives, which are necessary for the deformation of the considered IWs with any predetermined order of approximation $n$. The minimum number of neighboring points required to determine unknowns included in (3) is equal to $\sum_{m=0}^{n}(m+1)$, for example, at least 6 points are required to determine 6 unknowns for $2^{nd}$-order of accuracy. However, it is recommended to use a higher number of neighboring points to filter the errors embedded in the measured velocity vectors.

The first test case intended to qualitatively demonstrate the performance of the proposed method is a turbulent flow over a hemisphere mounted on a surface modeled by direct numerical simulation (DNS) (see figure 1a). A detailed description of this test case can be found in (Bobrov et al. 2021). The 8-bit PIV images with sizes of $L_x \times L_y = 1000 \times 200$ px, corresponding to a flow domain $x/D \times y/D = 5 \times 1$ located downstream of the hemisphere with diameter $D$ in the wall-normal plane, are generated using a conventional approach to PIV image generation (Lecordier and Westerweel 2004) with the following key parameters of the images. Particle images with the constant diameter of 3 px and the maximum value of individual particle image intensity $I_0 = 128$ are uniformly distributed within a light sheet with a Gaussian intensity profile and a thickness $\Delta Z/D = 0.04$, which results in averaged particle image concentration $N_I = 4$ per IW, representing low concentration. In order to simulate real experiments, Gaussian background noise with a root-mean-square of $\sigma_n = 0.2I_0$ and a mean value of $\mu_n = 5\sigma_n$ is superimposed on the PIV images. Three iterations are executed using an asymmetric shift of both IWs with respect to the corresponding mesh points (Wereley and Meinhart 2001), and a *sinc*-function with a kernel size of 8×8 px when interpolating the images. A universal median test (Westerweel and Scarano 2005) with standard parameters is applied after each iteration. For WIDIM, the outliers detected are replaced with a mean of the accepted neighboring values, and the velocity derivatives included in Eq. (1) are calculated using a central difference scheme. For the proposed method, the outliers detected are omitted and not taken into account when determining the velocity derivatives. When cross-correlating, FFT-based zero-normalized correlation is applied, which is computed with the help of an open-source FFTW library (Frigo and Johnson 2005). The resulting particle image displacements are defined with sub-pixel accuracy using a three-point parabolic peak fitting. The images are processed using the IWs of sizes $I \times I = 16 \times 16$ px with 75 % of overlap.

    Figures 1*b* and 1*c* demonstrate that increasing the order of the window deformation scheme in WIDIM does not always lead to better results. While in the case of the 1$^{st}$-order deformation, the relative number of outliers decreases with iterations from 6.9 % to 4.4 %, in the case of the 2$^{nd}$-order deformation, it increases dramatically to 27.3%. This loss of accuracy is associated with errors in the 2$^{nd}$-order derivatives and, as a result, with incorrect deformation of the IWs, leading to a low signal-to-noise ratio on the cross-correlation map. Figures 1*d* and 1*e* clearly demonstrate the advantages of the proposed method in terms of both the number of erroneous vectors and spatial resolution. Interestingly, the proposed method decreases the number of outliers down to 2.7 % and 4.0 % for the 1$^{st}$- and 2$^{nd}$-order deformation schemes, respectively, and accurately reconstructs them. Moreover, the small-scale flow structures become more pronounced (compare figure 1*e* with the exact flow field shown in figure 1*a*), demonstrating an increase in spatial resolution.

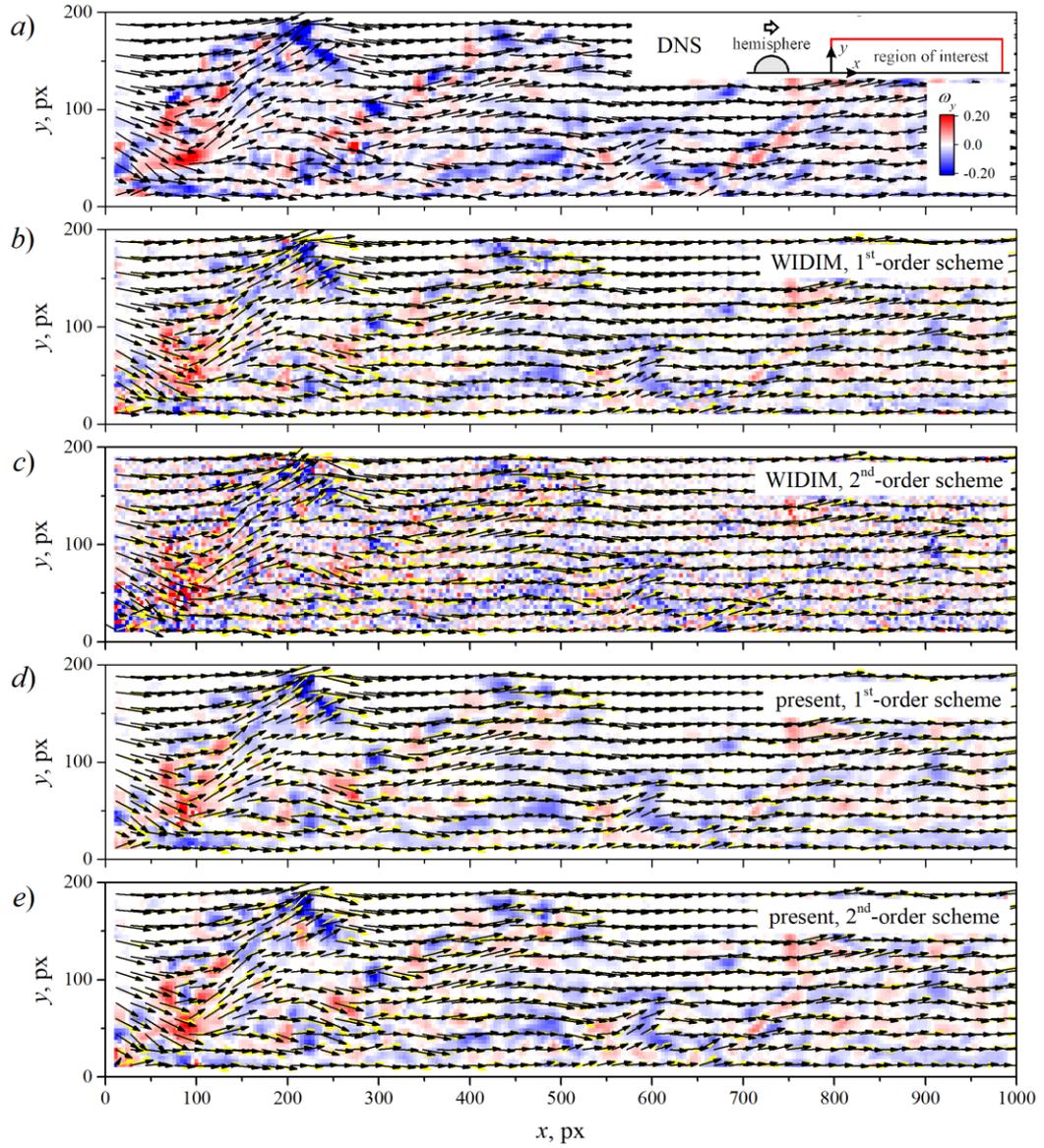

Figure 1. The instantaneous vorticity and vector velocity fields obtained by DNS (*a*), WIDIM (*b*, *c*), and proposed method (*d*, *e*) using 1$^{st}$- (*b*, *d*) and 2$^{nd}$- (*c*, *e*) order window-deformation schemes. For comparison, the exact velocity vectors colored in yellow are shown in (*b* - *e*).

In order to quantitatively estimate the performance of the proposed method, we consider the synthetically generated one-dimensional shearing displacement field, modeled as follows:

$$s = s_y = 3\sin\left[\frac{2\pi L_x}{\lambda_L - \lambda_0}\ln\left(1 + \frac{\lambda_L - \lambda_0}{L_x \lambda_0}x\right)\right]. \tag{6}$$

Eq. (6) takes into account the linear decrease in the wavenumber from $\lambda_0 = 256$ px at $x = 0$ to $\lambda_L = 16$ px at $x = L_x$ and allows one to estimate the performance of the proposed method in terms of spatial resolution. The generated dataset consists of 128 independent image pairs with sizes of $L_x \times L_y = 1000 \times 400$ px. Other image properties and processing parameters are the same as in the previous test case.

Figure 2 shows the results of this test case in terms of resolution $\lambda/I$ relative to IW size. A pronounced decrease in both random and bias errors is clearly seen in figure 2*a* for the results obtained by the proposed method, which yields several times smaller values of the random error down to the resolution $\lambda/I \approx 3$ in comparison with that obtained by WIDIM. The bias error is also reduced by the proposed method (see, for example, a two-fold reduction in the bias error at $\lambda/I \approx 2$). The reasons for these improvements lie in a more accurate determination of the derivatives. Figure 2b shows the estimates of the velocity derivative $ds/dx$, as the most challenging one compared to $ds/dy$. As it is seen, both methods yield almost the same levels of bias error. However, the random error for the proposed method is several times lower, up to the lowest considered resolution $\lambda/I \approx 1.2$. The same is applied to the $2^{nd}$-order derivative $d^2s/dx^2$ shown in figure 2c. Correct determination of all of these quantities with a low value of their rms has a positive effect on image deformation, leading to better particle image matching and a decrease in the number of outliers, as demonstrated in figure 2d. As a result, it leads to an increase in the accuracy of the measured displacements and spatial resolution (see figure 2a).

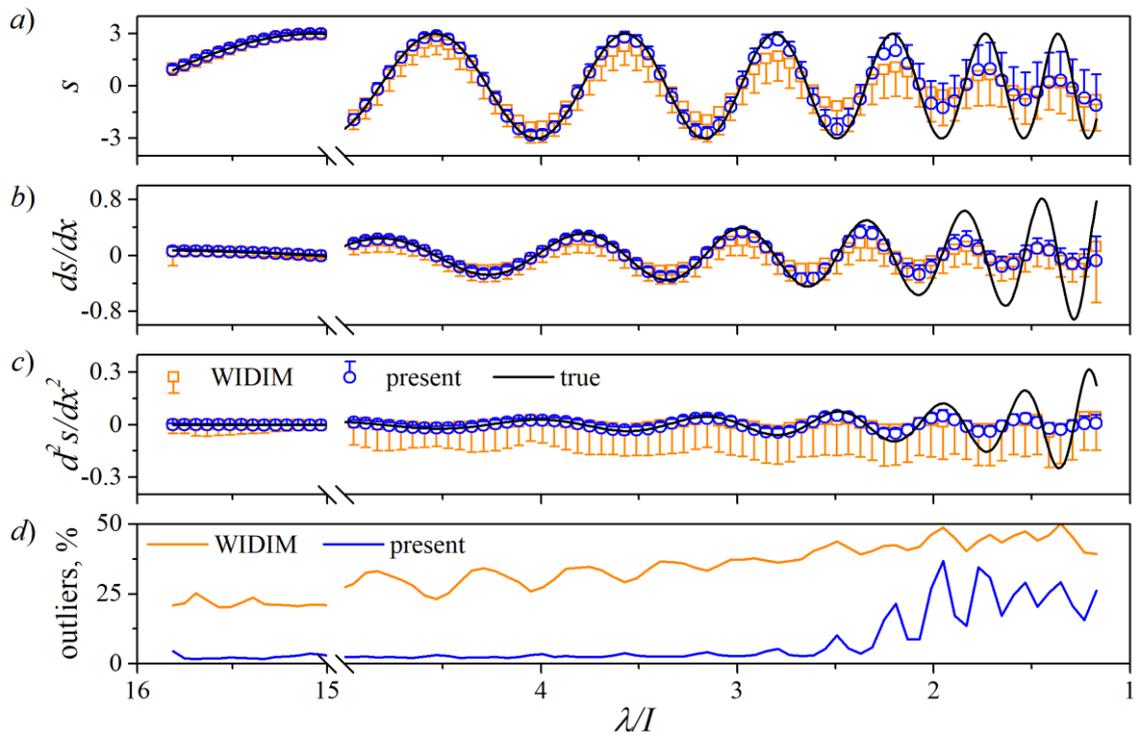

Figure 2. Exact (*solid black*) and measured (*symbols and bars corresponding to the mean and rms*) vertical particle image displacements (*a*) and their $1^{st}$- (*b*) and $2^{nd}$- (*c*) order derivatives in the *x*-direction. The bias error can be estimated by subtracting the exact value from the mean. Figure (*d*) represents the corresponding number of outliers arising after the third iteration.

In conclusion, the proposed method has shown a number of distinctive advantages. Due to its *grid-free* nature, it allows for the absence of some velocity vectors in the neighborhood of the mesh point under consideration, and, therefore, the step associated with the reconstruction of the missing velocity vectors is not required. This is useful at the boundaries of the computational domain, especially in the case of its complex geometry. The velocity derivatives are naturally obtained as a result of solving the system (2), avoiding the procedure of numerical calculation of the derivatives necessary for the IW deformation, and allowing direct determination of some flow quantities, such as vorticity. Further improvements may be associated with higher-order image deformation schemes taken into account in Eq. (3) and (4), and the selection of the optimal number of neighboring mesh points and matrix $W$ in Eq. (2).

This work was supported by the Russian Science Foundation (grant no. 22-29-01274).